\documentstyle[12pt]{article}
\begin{document}
\font\BF=cmbx10 scaled \magstep2
\begin{center}
{\BF
Gasdynamic Investigation of Rotating Gas Accretion}\\
G.~S.~Bisnovatyi-Kogan$^{*}$ and N.~V.~Pogorelov$^{**}$\\
\medskip
$^{*}$Space Research Institute, Russian Academy of
Sciences,\\
84/32\ Profsoyuznaya St., 117810 Moscow, Russia\\
\medskip
$^{**}$Institute for Problems in Mechanics, Russian Academy of
Sciences,\\
101\ Verndskii Ave.,
Moscow 117526, Russia
\end{center}
\vskip 2.truecm
%\vfill \eject

\begin{abstract}
Gasdynamic features of slowly rotating gas axisymmetric accretion onto
a gravitating center are investigated. The process of flow restructuring
is studied as the angular velocity of accreting matter approaches
the Keplerian angular velocity. For spherically-symmetric accretion, the
conditions are found of the existence of a steady-state supersonic solution
for various external boundary conditions. The numerical study is performed on
the basis of the Lax--Friedrichs-type second order of accuracy
high-resolution numerical scheme with the implicit approximation
of the source term in the Euler equations.
\end{abstract}

\section{Introduction.}
Accretion onto neutron stars and black holes gives the main energy supply in
galactic X-ray sources. Among X-ray sources with high mass companions there
are long-periodic pulsars, whose origin is not yet clear~
(Nagase~1989; Lipunov~1992). Owing to high-speed stellar wind, the angular
momentum of the falling matter is often not sufficient for the accretion disk
formation at the level of the Alfv\'enic radius where the magnetic pressure
of the neutron star is balanced by the dynamic pressure of the falling gas.
The angular momentum acquired in this case by the neutron
star from the falling gas was shown (Bisnovatyi-Kogan~1991) to be such that
the equilibrium rotational period of the X-ray pulsar might be large. It is
rather difficult to make estimations for the equilibrium rotational period
because the formation is possible of outflowing streams which carry away the
angular momentum (Illarionov \& Kompaneetz~1990; Lovelace~1995). The
accretion picture in this case is three-dimensional and its numerical
modeling is very complicated. It was performed only for the case of accretion
onto a gravitating center, showing the presence of the Rayleigh--Taylor
instabilities (Matsuda et al.~1989). The nature of these instabilities is
unclear and their numerical origin can not be excluded (Steinolfson et
al.~1994).

To investigate the formation of long-periodic pulsars, the model is necessary
for the accretion onto a magnetized neutron star from the stellar wind in the
binary. To reduce a 3D problem to two dimensions, we can consider
either the conical accretion of nonrotating gas or the accretion of slowly
rotating gas onto a stationary star. In the first case (Koide et al.~1991),
the average angular momentum acquired by the neutron star is zero due to the
symmetry of the flow, so it can not be applied to long-periodic X-ray
pulsars.

Here we use the second approach. Our aim is to consider accretion onto a
magnetized star with the Alfv\'enic radius $R_A \gg R_\ast$~(star radius)
taking into account the interaction between the gas and the magnetosphere and
the possible outflow formation. As the first step, we consider accretion with
the full penetration of the gas through the magnetosphere substituted by the
inner boundary with the radius $R_\ast$ and neglect magnetogasdynamic
effects.

Beskin \&~Pidoprygora~(1995) presented the approximate solution for the
accretion of nonrotating gas onto a slowly rotating black hole in the
framework of the general theory of relativity. In this work, we consider the
accretion of rotating gas in the Newton approximation up to the rotation
velocities close to the Keplerian velocity.

Three different modes are usually considered as constituent parts of the
astrophysical accretion and are fairly well investigated (Lipunov~1992;
Bisnovatyi-Kogan~1989).
\begin{enumerate}
\item Spherically-symmetric accretion occurs if
 the star velocity $v_\infty$ is much smaller than the speed of sound
 $a_\infty$ in an accreting matter and the angular momentum is negligible.
 \item Cylindrical accretion is realized when $v_{\infty} \ge a_{\infty}$ with
 vanishing angular momentum.
 \item Accretion disk is formed if the total angular momentum of the matter
 is sufficient for its formation.
\end{enumerate}

Real accretion is, in fact, a combination of the above-mentioned modes.

From gasdynamic viewpoint, it is of interest to investigate the process of
the transition from regime 1 to regime 3 for $v_\infty \ll a_\infty$ as the
rotational velocity of the accreting matter approaches the Keplerian
velocity. We consider the polytropic flow of ideal, perfect gas with the
polytropic index $\gamma=1.4$.  Calculations are performed using the second
order of accuracy in space high-resolution Lax--Friedrichs-type numerical
scheme proposed by one of the authors~(Barmin \&~Pogorelov~1995). Detailed
description of the scheme is given by Barmin, Kulikovskiy \&~
Pogorelov~(1996).  Recently, a three-dimensional Bondi--Hoyle accretion
was investigated for the accretor~(star) radius varying from 10 to 0.02 Bondi
radii, so that both subsonic and supersonic regimes of accretion could be
realized~(Ruffert~1994). In our study we are mainly interested in qualitative
tracing the flow restructuring as its angular velocity increases. For this
reason, the Euler equations are solved and the external flow is supposed to
be supersonic. On the other hand, the size of the star is assumed to be
sufficiently small; the inner, also supersonic, boundary is fixed at a finite
distance from the gravitational center. The effect of the external boundary
conditions is investigated on the existence and properties of a stationary
supersonic accretion for the chosen computational region.

\section{Spherically-symmetric supersonic accretion.}

Let us consider the accretion gas flow between two spheres with the inner and
outer radii equal to $R_\ast$ and $R_0$ respectively. If the flow at the
distance $R_0$ is assumed to be supersonic, then all flow parameters should
be fixed at this boundary. The question is what parameter values can be
imposed at a given distance for the existence of a stationary solution. If
the flow is shockless and polytropic,
the following three
conservation equations must be satisfied:
\begin{eqnarray}
 & &4 \pi R^2 \rho U = \dot M, \\
 & & \nonumber \\
 & & {p \over \rho^\gamma} = K, \\
 & & \nonumber \\
 & & {U^2 \over 2} +{\gamma \over \gamma -1} {p \over \rho} - {G M \over R} =
     h_{t0}
\end{eqnarray}

Here $\rho$, $p$, $U$ are, respectively, the density, the pressure and the
radial velocity, $R$ is the current distance from the gravitating center, $M$
is the mass of the star.  The values of the accretion rate $\dot M$,
the constant $K$, and the Bernoulli constant $h_{t0}$ are to be fixed at $R =
R_0$.  Introducing dimensionless variables with the units of velocity,
pressure, density, and length equal to $U_0$, $\rho_0 v_0^2$, $\rho_0$, and
$R_\ast$, respectively, and designating $M_0 = U_0 / a_0$, $a_0 =(\gamma p_0
/ \rho_0)^{1/2}$ and $S = G M / U_0^{2} R_\ast$, we can rewrite
system~(1)--(3) in the dimensionless form using the same notations
for nondimensional values of $U$, $R$, $p$, $\rho$, and $a=(\gamma
p/\rho)^{1/2}$ (indices ``0'' and ``$\ast$'' correspond to the outer and to
the inner boundary, respectively):

\begin{eqnarray}
 {U^2 \over 2} + {a^2 \over \gamma -1}-{S \over R}&=& {1
 \over 2} + {1 \over (\gamma -1) M_0^2}-{S \over R_0}, \\
 & & \nonumber \\
 \rho U&=&{R_0^2 \over R^2}, \\
 & & \nonumber \\
 {p \over \rho^\gamma}&=&{1 \over \gamma M_0^2}
\end{eqnarray}

Being subsonic at infinity and supersonic at $R = R_0$, the flow becomes
sonic $U=U_B=a_B$ at some point $R = R_B\ge R_\ast$,
where~(Bisnovatyi-Kogan~1989)
\begin{equation}
U_B=a_B ={1 \over 2}{S \over R_B}
\end{equation}

In the sonic point Eqs.~(4)--(7) reduce to
\begin{equation}
   {5-3\gamma \over 4(\gamma-1)} {S \over R_B} = h_{t0}
\end{equation}

If $R_B \ge R_0$ and the flow comes from infinity, the value $h_{t0}$ must
satisfy the inequality
\begin{equation}
  0 < h_{t0} < {5-3\gamma \over 4 (\gamma-1) R_0} S
\end{equation}

\section{Accretion with rotation.}
In this section we consider the process of the axisymmetric rotating flow
accretion onto a gravitating object. Analysis is performed on the basis of
the numerical solution of the Euler gasdynamic equations.  The gas is 
assumed
to be ideal and perfect with the following caloric equation of state:
$\varepsilon=p/(\gamma -1) \rho$, where $\varepsilon$ is the internal energy
per unit mass and $\gamma=1.4$ is the polytropic index.  The system of
governing equations in the Cartesian coordinate system shown in Fig.~1 reads:
\begin{equation}
 {\partial {\bf U}\over \partial t} +
 {\partial {\bf E}\over \partial x} + {\partial {\bf G}\over \partial z} +
  {\bf H} = 0,
\end{equation}
where
\[
    \def\quad{\hskip.5em}
    {\bf U} = \left[\matrix{
    \rho\cr \rho u\cr \rho v\cr \rho w\cr e\cr}
    \right]\!,\quad
    {\bf E}= \left[\matrix{
      \rho u\cr \rho u^{2}+ p\cr \rho uv\cr \rho uw\cr (e+p)u\cr}
    \right]\!,\quad
    {\bf G} = \left[\matrix{
      \rho w\cr \rho uw\cr \rho vw\cr \rho w^{2}+p\cr (e+p)w\cr}
    \right]\!,\quad
    {\bf H} = {\rho \over x}\left[\matrix{
       u\cr
       \displaystyle (u^2-v^2)+GMx{\sin \theta \over R^2}\cr 2uv\cr
       \displaystyle uw + GMx{\cos \theta \over R^2}\cr
       \displaystyle {(e+p)u \over \rho}+ 
       GMx{u \sin \theta  + w \cos \theta\over R^2}\cr} \right]\!
\]

Here $e= p/(\gamma-1) + \rho (u^2+v^2+w^2)/2$ is the total energy
per unit volume. As far as the gravitational field is considered
spherically symmetric, the polar computational domain $(R,\theta)$ is
chosen with the inner and outer radii equal to $R_\ast$ and $R_0$,
see Fig.~1. On normalizing the quantities of  density, pressure, and
velocity by $\rho_0$, $\omega_{\ast} R_{\ast}$, $\rho_0 \omega^2_\ast
R^2_\ast$, where $\rho_0$ is the density at $R = R_0$ and $\omega_\ast$ is
the gas angular velocity at $R = R_\ast$ at the equator for
the constant angular momentum distribution,
the source term can be rewritten in the dimensionless form
as
\begin{equation}
    {\bf H} = {\rho \over x}\left[
    \begin{array}{c}
       u\\
       \displaystyle(u^2-v^2)+Sx{{\sin \theta} \over {R^2}}\\
       2uv\\
       \displaystyle uw + Sx{\cos \theta \over R^2}\\
       \displaystyle
       {(e+p)u \over \rho} + Sx{u \sin \theta  + w \cos \theta \over R^2}
\end{array}
    \right]
\end{equation}

Here $S=GM/{\omega^2_\ast R_\ast^3}$.  The form of the other vector
components in Eq.~(10) remains unchanged.

From now on we consider only nondimensional parameters.

The following procedure is used to construct the initial and boundary
conditions.

1. Introducing dimensionless parameter $\alpha = U_0/U_{K\ast}$, where $U_0$
is the radial velocity on the outer boundary and $\ U_{K\ast} =
(GM/R_\ast)^{1/2}$ is the Keplerian velocity on the inner boundary, we fix
the values on the outer boundary as if the flow is spherically symmetric:
\[
  \rho_0 = 1;\ U_0 = \alpha S^{1/2};\ p_0 = \alpha^2 S/\gamma M_0^2,\
  W_0=0,
\]
where $W_0$ is the $\theta$-component of the velocity and $M_0=U_0/a_0$.

Corresponding parameter values inside the computational region are adopted
equal to those at the boundary.

We can find the entropy and the total enthalpy dimensionless values as
follows:
\[
  {p_0 \over \rho_0^\gamma}=K; \qquad h_{t0}=S \left[\alpha^2
  \left({1 \over (\gamma-1)M^2_0} + {1 \over 2}\right) -{1 \over R_0}\right]
\]

2. The following initial distribution of the angular velocity is assumed:
\[
  \left\{
  \begin{array}{ll}
  \mbox{$\displaystyle\omega = {1 \over x^2} = {1 \over R^2 {\sin^2 \theta}}$}
  \hskip 2truecm
  & \mbox{if $R>20$~(constant angular momentum)}\\
  \mbox{$\omega = 0.0025$}
  \hskip 2truecm
  &\mbox{if $R \leq 20$~(constant angular velocity)}
  \end{array}
  \right.
\]

The $y$-component $v$~(normal to the plane of Fig.~1) of the velocity $\bf v$
is then $v=\omega x$.

3. The values of $\rho$ and $p$ in the whole computational region are
modified. Assuming $U(R,\theta) = U_0$ and $W(R,\theta)=0$, we find the new
pressure and density values from the formulas
\[
  {\gamma \over \gamma -1}{p \over \rho} + {U^2 \over 2} +
  {v^2 \over 2} - {S \over R} = h_{t0}; \qquad {p \over \rho^\gamma} = K
\]

After that all values on the external boundary are fixed because it is
supersonic. No boundary conditions are necessary on the inner circle, since
the flow through it is supersonic.

\section{Numerical scheme.}
Point clustering towards the internal boundary circle is made to obtain a
sufficiently fine flow resolution in the vicinity of the gravitating center.
The following formula is used:
\[
  R = R_{\ast} + \left(R_0 - R_{\ast}\right){e^{\beta \xi} -1 \over
  e^{\beta} -1}
\]
with the clustering parameter $\beta$.

If we introduce a polar mesh
\begin{eqnarray*}
    & &\xi_{l} = (l-1)\Delta \xi ,\; l=1,2,\ldots,L;\;
    \theta _{n} = (n-2.5) \Delta \theta,\; n=1,2,\ldots,N;\\
    & &R_l = R(\xi_l),\; \Delta \xi = 1 / (L-1),\;
    \Delta \theta = \pi /(2 N - 8)
\end{eqnarray*}
with the center in the accretor position, then for each cell the system~(1)
in the finite-volume formulation acquires the form
\begin{eqnarray}
 & &R_{l}\,\Delta R_l\,\Delta\theta\,{\partial{\bf U}^k_{l,n}\over\partial t}
 +{} \, (R_{l+1/2}\bar {\bf E}^k_{l+1/2,n} + R_{l-1/2}\bar {\bf
    E}^k_{l-1/2,n}) \Delta \theta+{} \nonumber \\
 & & (\bar {\bf E}^k_{l,n+1/2} +\bar {\bf E}^k_{l,n-1/2})\, \Delta R_l + {}\,
     R_{l}\, \Delta R_l\, \Delta \theta\,  {\bf H}^{k+1}_{l,n} = 0
\end{eqnarray}

Here $\Delta R_l = R_{l+1/2} - R_{l-1/2}$ and
$\bar{\bf E}$ is the  flux normal to the boundary, defined as:
$$
\bar{\bf E} = n_1 {\bf E} + n_2{\bf G},
$$
where ${\bf n}= (n_{1},n_{2})$  is a unit outward vector normal to the cell
surface.

The assumption is made of a piecewise-parabolic distribution  of  the
primitive gasdynamic parameters $\bf q$ inside the cells to specify values on
their boundaries and  slope delimiters are used to obtain the nonoscillatory
property.

The averaged slope inside the cell in the radial direction is determined as
~(Sawada et al.~1989)
\[
 \Delta q_l = [(2 \Delta R_{l-1} + \Delta R_l) \mu\, + \,
 (2 \Delta R_{l+1} + \Delta R_l) \nu] \Delta R_l\ /\chi,
\]
where
\[
\chi = \Delta R_{l+1} + \Delta R_l +  \Delta R_{l-1},\;
\mu = {\delta q_{l+1/2} \over \Delta R_{l+1} + \Delta R_l},\;
\nu = {\delta q_{l-1/2} \over \Delta R_l + \Delta R_{l-1}}
\]

The left and the right boundary value are then estimated as
\begin{eqnarray*}
  & & q_{l+1/2}^L = q_l + \frac{1}{2}\, \tilde \Delta q_l,\
      q_{l-1/2}^R = q_l - \frac{1}{2}\, \tilde \Delta q_l,\\
  & & \tilde \Delta q_l = {\rm minmod}(\Delta q_l, 2 \mu \Delta R_l,
     2 \nu \Delta R_l)
\end{eqnarray*}

Here
\[
  \mbox{minmod$(a,b,c)$}= \left\{
  \begin{array}{ll}
  \mbox{sign$(a) \min (|a|, |b|, |c|)$} \quad & \mbox{if $b c > 0$}  \\
  0 & \mbox{otherwise}
  \end{array}
  \right.
\]

Found the values of ${\bf q}^L$ and ${\bf q}^R$ on the both sides of the cell
boundary, the appropriate fluxes $\bar {\bf E}$ are calculated using the
modified Lax--Friedrichs formula~(Barmin \&~Pogorelov~1995):
\[ \bar{\bf E}({\bf U}^R,{\bf
  U}^L) = \frac{1}{2} [\bar{\bf E}({\bf U}^L) + \bar{\bf E}({\bf U}^R) - \hat
  {\bf R} ({\bf U}^R - {\bf U}^L)],
\]
where the matrix $\hat {\bf R}$ is a positive diagonal matrix with the
entries equal to the spectral radius (the maximum of eigenvalue 
magnitudes) of ${\partial \bar{\bf E} \over \partial {\bf U}}$.  This 
scheme gives a drastic
simplification of the numerical algorithm comparing to the methods based on
precise characteristic splitting of the Jacobian matrices, while preserving
nonoscillatory property.  Having the second order of accuracy, it is less
dissipative than the original Lax--Friedrichs scheme.  Similarly, the fluxes
through another pair  of cell  surfaces are obtained.

As seen from Eq. (12) the source term is approximated implicitly for
better stability of the numerical scheme. A proper linearization of this term
is made to realize the numerical procedure of obtaining the
time-converging steady-state solution:
\[
  {\bf H}^{k+1}= {\bf H}^k + {\partial {\bf H}^k \over \partial {\bf U}^k}
  \ ({\bf U}^{k+1} - {\bf U}^k).
\]
The promotion of the solution in time is  performed  with the first order of
accuracy by the formula:
\begin{eqnarray*}
& &(I\ +\ \Delta t {\partial {\bf H}^k \over \partial {\bf U}^k}) 
   ({\bf U}^{k+1}_{l,n}
 - {\bf U}^k_{l,n}) = - \Delta t \bigl[ 
   ( R_{l+1/2}\bar {\bf E}^k_{l+1/2,n} / R_l
  + R_{l-1/2}\bar {\bf E}^k_{l-1/2,n}/ R_l) / {\Delta R}_l + \nonumber \\
& & (\bar {\bf E}^k_{l,n+1/2} +\bar {\bf E}^k_{l,n-1/2})\,/ R_l \Delta \theta
  +  {\bf H}^k_{l,n} \bigr].
\end{eqnarray*}
for $t = k\, \Delta t$, $k=0,1,\ldots\,$, and $\Delta t$ defined by
the CFL condition ($I$ is the identity matrix).

\section{Analysis of numerical results.}

All results presented in this section were obtained in the ring region with
the dimensionless inner and outer circle radii being $R_\ast=1$ and
$R_0=100$, with 56 cells in the angular direction and 104 cells in the radial
direction, and with the clustering parameter $\beta =4$. The calculations
were performed until their full time-convergence in a quarter of the ring
with the appropriate reflection conditions applied in the planes of the flow
symmetry.

The way of presentation the results obtained is the following. In Figs.~2--18
the isolines of different gasdynamic parameters and the streamlines of the
flow are presented in the lower and upper parts of the figure divided by the
rotational axis.  Figures~2--3, 7--9, and~13--15 correspond to the whole
computational region. In these figures 18 isolines are presented with the
constant step between the maximum and the minimum value of functions
indicated in the corners, that is, the isoline value for any function $f$ can
be found by the formula $f_i=f_{\rm min}+i \times (f_{\rm max}-f_{\rm
min})/19$.  In the regions with no captions in the corner, the streamlines
are shown.  Figures~4--6, 10--12, and~16--18 present the
magnified central part (50 computational zones) of the corresponding figures
related to the whole computational region.

The following dimensionless parameters are chosen, at first, to study the
accretion flow behavior for the case of slow rotation~(this choice is
consistent with the considerations from Section 2):
\[
 \alpha=0.1;\  \gamma=1.4;\ M_0=2;\ S=250
\]

The steady state results for this case are presented in Figs.~2--6 (the axis
of rotation is aligned with the $z$-axis~(see Fig.~1). In Fig.~2, the
isolines of the pressure and the density
decimal logarithm are presented
above and below the rotational axis, respectively, in the whole computational
region.  The isolines of the velocity components $v$ and
$W$($\theta$-direction) are not shown in the full computational region
since their main variation is located in the vicinity of the inner boundary.
The isolines of the velocity component $U$ (radial direction) and the
streamlines are given in Fig.~3.

In Figs.~4--6 the same lines are shown in the magnified subregion close to the
accreting center (50 computational zones). It is clearly seen in these
figures that for long distances from the accreting center and far enough from
the rotational axis slow rotation does not affect the flow significantly and
it is, in fact, the superposition of a spherically symmetric accretion and an
axisymmetric constant angular momentum rotation. Closer to the $z$-axis
and to the accreting center, however, the pressure and the density are
greater at the equator than near the poles. The deviation from the purely
constant angular momentum rotation appears only in the vicinity of
the rotational axis and near the accreting object. The streamlines and
the radial velocity $U$ isolines behave quite like in the spherically
symmetric case.

The similar pictures of the flow are presented, for $S=25$ with the rest
of dimensionless parameters unchanged, in Figs.~7--12. The effect of rotation
in this case is quite definite in the whole region surrounding the rotational
axis. The gas displacement from the poles is much more pronounced comparing
with the variant of $S=250$ even at large distances from the accretion
center, see Fig.~7.  The streamlines deflect from the poles to the equator
region~(Fig.~12) and the size of the domain with substantial values of
$W$ is larger (Figs.~8,~11).

The picture of the flow for the rotational velocity close to the Keplerian
velocity ($S=1.5$) is shown in Figs.~13--18. Almost all the gas is
removed in this case from the pole regions. The density near the equator is
$\sim 10^4$ times greater than that near the poles and almost all accreting
matter is involved in the motion towards $\theta = \pi /2$. The structure of
the flow is very close to the picture of the accretion disk formation, see
Fig.~18 showing the streamline distribution. The regions with the dense
distribution of the pressure and density isolines~(Fig.~16) represent the
oblique transverse shock where supersonic flow of the gas in
$\theta$-direction, declined by the centrifugal force from the poles to the
equator, decelerates to become zero at $\theta = \pi /2$.  Behind these shocks
the region of a fast rotation is observed~(Figs.~14,~17). This region of a
highly rotating dense matter will form an accretion disk if $S < 1$.

\section{Discussion.}

Numerical modeling is performed for axisymmetric rotating gas accretion
onto the star. The second order of accuracy in space
high-resolution Lax--Friedrichs-type scheme showed good performance for the
calculation of flows with an extremely great variation of the pressure
and density values throughout the computational region.
The flow restructuring is investigated as the rotation velocity
approaches the Keplerian velocity. The variants are presented for
different cases begining from a slow rotation and up to the case,
for which the steady state centrifugal and gravitational forces near the
stellar equator are rather close. The values of the angular momentum,
however, were sufficiently small, so that at the inner boundary the
centrifugal force was smaller than gravitational, and the stationary accretion
picture could be attained. If the angular momentum of the falling matter
reaches the value, for which the centrifugal force balances the gravitation
before the matter falls onto the star (the inner boundary in our
calculations), the matter stops near the equatorial plane and forms a disk.
In the absence of viscosity the mass of this disk increases
in time.  In the presence of viscosity, which is especially efficient in
the turbulent case, the matter in the disk looses its angular momentum and
moves slowly towards the center forming a stationary accretion
disk~(Lynden-Bell~1969; Shakura~1972).  We are interested here in the problem
of the long-periodic pulsar formation, for which the angular momentum is not
sufficient for the accretion disk origin near the Alfv\'enic surface where
the dynamic pressure of the flow is balanced by the magnetic surface at the
magnetosphere of the neutron star. To find the angular momentum of the matter
falling onto the star during accretion from the stellar wind penetrating
through the Alfv\'enic surface, we need to include into consideration the
magnetic field.  This work is now in progress.

\bigskip
{\bf Acknowledgement.}
This work was supported by the Russian Foundation of Basic Research under
Grant No.~95-01-00835.

%{\footnotesize

\vfill\eject

\centerline{\bf Figure captions.}

\medskip
Fig.~1. Computational region.

Fig.~2. Logarithm pressure and density isolines.
               Full computational region, $S=250$.

Fig.~3. Streamlines and $U$ isolines.
               Full computational region, $S=250$.

Fig.~4. Logarithm pressure and density isolines.
               Inner subregion, $S=250$.

Fig.~5. Isolines of $v$ and $W$. Inner subregion, $S=250$.

Fig.~6. Streamlines and $U$ isolines. Inner subregion,
               $S=250$.

Fig.~7. Logarithm pressure and density isolines.
               Full computational region, $S=25$.

Fig.~8. Isolines of $v$ and $W$.
               Full computational region, $S=25$.

Fig.~9. Streamlines and $U$ isolines.
               Full computational region, $S=25$.

Fig.~10. Logarithm pressure and density isolines.
               Inner subregion, $S=25$.

Fig.~11. Isolines of $v$ and $W$.
               Inner subregion, $S=25$.

Fig.~12. Streamlines and $U$ isolines.
               Inner subregion, $S=25$.

Fig.~13. Logarithm pressure and density isolines.
               Full computational region, $S=1.5$.

Fig.~14. Isolines of $v$ and $W$.
               Full computational region, $S=1.5$.

Fig.~15. Streamlines and $U$ isolines.
               Full computational region, $S=1.5$.

Fig.~16. Logarithm pressure and density isolines.
               Inner subregion, $S=1.5$.

Fig.~17. Isolines of $v$ and $W$.
               Inner subregion, $S=1.5$.

Fig.~18. Streamlines and $U$ isolines.
               Inner subregion, $S=1.5$.

\end{document}